\documentclass[conference]{IEEEtran}
\IEEEoverridecommandlockouts
\usepackage{cite}
\usepackage{amsmath,amssymb,amsfonts}
\usepackage{mathtools}
\usepackage{graphicx}
\usepackage{textcomp}
\usepackage{xcolor}
\usepackage{braket}
\usepackage[caption=false,font=footnotesize]{subfig}
\usepackage{hyperref}
\usepackage{enumitem}
\usepackage{placeins}
\usepackage{algorithm}
\usepackage{algpseudocode}
\usepackage{bm}% bold math
\usepackage{lipsum}

\newcommand{\ir}{\Lambda_{\mathrm{IR}}}
\newcommand{\uv}{\Lambda_{\mathrm{UV}}}

\begin{document}

\title{Parameter Calibration for Reduced-Bandwidth Two-Photon Waveguide-QED Simulations}

\author{
Romain Piron\IEEEauthorrefmark{1} and Akihito Soeda\IEEEauthorrefmark{1}\IEEEauthorrefmark{2}\IEEEauthorrefmark{3}
\\[0.6em]
\IEEEauthorrefmark{1}\textit{Principles of Informatics Research Division, National Institute of Informatics,} \\
\textit{2-1-2 Hitotsubashi, Chiyoda-ku, Tokyo, Japan} \\
\IEEEauthorrefmark{2} \textit{Department of Informatics, School of Multidisciplinary Sciences, SOKENDAI, Japan} \\
\IEEEauthorrefmark{3} \textit{Department of Physics, Graduate School of Science, The University of Tokyo, Japan} \\[0.6em]
\texttt{\small \{piron,soeda\}@nii.ac.jp}
}

\maketitle

\begin{abstract}
     Waveguide-QED platforms represent one potential approach to scalable quantum technologies, but their simulation remains computationally demanding due to the large number of frequency modes required to describe traveling photons. In practice, increasing the simulated bandwidth rapidly raises the numerical cost, leading to a trade-off between accuracy and tractability. The existing approaches formulated in time-domain indirectly control this trade-off through the choice of time step, which obscures the connection between discretization parameters and the represented spectral window. In this work, we introduce an end-to-end framework to explicitly control the effective bandwidth in waveguide-QED simulations of two-photon scattering. We show that truncating the frequency domain requires consistent shifts of the model parameters, and derive a systematic calibration procedure that preserves the physical accuracy of the reduced model. This enables tuning the central frequency and the bandwidth of the numerical spectrum, leading to a several-fold reduction in the Hilbert space dimension while maintaining physical accuracy. We discuss the limitations of this calibration and relate the finite-bandwidth viewpoint to time-domain discretizations.
\end{abstract}

\begin{IEEEkeywords}
Quantum Optics, Optical Quantum Computing, Quantum Networking, Numerical Simulation, Waveguide-QED
\end{IEEEkeywords}

\section{Introduction}

Photonic technologies for generating and controlling traveling photons are expected to play a central role in quantum information processing. In this context, waveguide quantum electrodynamics (QED) has emerged as a versatile platform to engineer strong and controllable light-matter interactions. In such systems, fixed emitters interact via the exchange of photons confined to one-dimensional waveguides, giving rise to collective many-body effects with nontrivial quantum correlations relevant for quantum information tasks~\cite{royStronglyInteractingPhotons2017, sheremetWaveguideQuantumElectrodynamics2022,gonzalez-tudelaLightMatterInteractions2024}. Beyond emitter-based quantum control, waveguide-QED systems also provide a natural route to induce effective photon-photon interactions: while photons interact only weakly in vacuum -- motivating linear-optics quantum computing schemes such as the KLM proposal~\cite{knillSchemeEfficientQuantum2001} -- their coupling to quantum emitters enables strong effective photon-photon interactions~\cite{royStronglyInteractingPhotons2017}. Finally, the interplay between localized emitters and propagating photons makes waveguide-QED systems a promising platform for quantum networking, where emitters can act as effective linear optical elements in the monochromatic limit~\cite{rouletTwoPhotonsAtomic2016}, and where recent nanofiber-based architectures highlight their potential for scalable quantum networks and enhanced interaction lengths in quantum memories~\cite{sheremetWaveguideQuantumElectrodynamics2022,gonzalez-tudelaLightMatterInteractions2024, sunamiScalableNetworkingNeutralAtom2025}.

Traveling photons in waveguide-QED systems have been extensively studied in the literature. Early works~\cite{shenCoherentSinglePhoton2005} introduced Hamiltonian descriptions of single-photon transport interacting with a two-level system (TLS), later extended to multiple photons~\cite{shenStronglyCorrelatedTwoPhoton2007,shenStronglycorrelatedMultiparticleTransport2007} and multiple emitters~\cite{shiMultiphotonScatteringTheory2015a}. Comprehensive reviews~\cite{sheremetWaveguideQuantumElectrodynamics2022,royStronglyInteractingPhotons2017}, along with more recent theoretical developments~\cite{greenbergSinglephotonScatteringQubit2023,pletyukhovScatteringMasslessParticles2012, schneiderGreensFunctionFormalism2016a}, provide a broad overview of these models and the analytical techniques used to study them. However, accurately describing photon propagation requires accounting for a large number of electromagnetic modes, making analytical treatments involved and typically reliant on restrictive assumptions that limit their applicability to specific regimes~\cite{sheremetWaveguideQuantumElectrodynamics2022,royStronglyInteractingPhotons2017}. This motivates the development of scalable numerical approaches capable of faithfully emulating waveguide-QED dynamics in regimes relevant for quantum technologies~\cite{brown5YearUpdate2024}.

A few numerical methods are currently available to simulate traveling photons in waveguide-QED systems (see Table 1 of Ref.~\cite{bundgaard-nielsenWaveguideQEDjlEfficientFramework2025}). Recent developments include the Julia package \texttt{WaveguideQED.jl}~\cite{bundgaard-nielsenWaveguideQEDjlEfficientFramework2025}, which efficiently simulates photon scattering from a localized emitter using a time-bin discretization of the wavepacket. Other approaches rely on matrix product states (MPS) or spatially discretized waveguide models~\cite{arranzregidorModelingQuantumLightmatter2021}, as implemented for instance in the Python package \texttt{QwaveMPS}~\cite{regidorQwaveMPSEfficientOpensource2026}, or more generally on tensor-network techniques~\cite{khanTensorNetworkApproach2025}. Alternative frameworks are based on direct implementations of input-output relations~\cite{fanInputOutputFormalismFewPhoton2010,kiilerichInputOutputTheoryQuantum2019}. Despite their methodological differences, these approaches share a common feature: they rely on a transformation to the time domain of the photonic operators. While this representation provides a convenient description of the interaction Hamiltonian for Markovian emitters~\cite{bundgaard-nielsenWaveguideQEDjlEfficientFramework2025,arranzregidorModelingQuantumLightmatter2021,regidorQwaveMPSEfficientOpensource2026,khanTensorNetworkApproach2025,fanInputOutputFormalismFewPhoton2010,kiilerichInputOutputTheoryQuantum2019}, it is less naturally suited to systems involving non-flat spectral densities. For instance, within such time-domain formulations, simulating non-Markovian effects require the explicit introduction of feedback loops~\cite{bundgaard-nielsenWaveguideQEDjlEfficientFramework2025,arranzregidorModelingQuantumLightmatter2021}.

This motivates the continued investigation of complementary numerical approaches formulated directly in the frequency domain, building on early numerical studies of light-matter interactions. In this representation, the Hamiltonian is defined in terms of frequency photonic modes interacting with localized emitters~\cite{scullyQuantumOptics1997,havukainenQuantumSimulationsOptical1999}. This approach also remains central to many analytical techniques, including functional methods in scattering theory~\cite{schneiderGreensFunctionFormalism2016a,pletyukhovScatteringMasslessParticles2012}. From a numerical perspective, working in the frequency domain provides a direct and physically transparent control of the Hilbert space dimension, which is set by the number of frequency modes included in the model~\cite{obaFastSimulationMultiphoton2024,pironRenormalizationTreatmentIR2026}. In contrast, time-domain approaches relate the effective system size to the time discretization step~\cite{bundgaard-nielsenWaveguideQEDjlEfficientFramework2025,arranzregidorModelingQuantumLightmatter2021,regidorQwaveMPSEfficientOpensource2026, khanTensorNetworkApproach2025,fanInputOutputFormalismFewPhoton2010,kiilerichInputOutputTheoryQuantum2019}, whose connection to the underlying energy scales is less direct, making control of the Hilbert space dimension with respect to these energies less transparent.

While the frequency-domain representation makes the spectral content of the numerical model explicit, truncating the photonic spectrum modifies the physical behavior of the simulation, even when the retained window contains the relevant modes. This effect can be understood within the framework of renormalization theory, where restricting the spectral bandwidth induces shifts in the effective parameters of the system~\cite{peskinIntroductionQuantumField1995}. Such effects have been identified in waveguide-QED settings~\cite{schneiderGreensFunctionFormalism2016a,sheremetWaveguideQuantumElectrodynamics2022,greenbergSinglephotonScatteringQubit2023}, and recent work suggests that they can be consistently compensated~\cite{pironRenormalizationTreatmentIR2026}, enabling reduced-bandwidth simulations that preserve physical accuracy. However, the practical relevance of this single-excitation calibration beyond the single-photon setting remains to be assessed, especially in explicit two-photon simulations where reducing the number of retained modes becomes substantially more important.

In this work, we build upon Ref.~\cite{pironRenormalizationTreatmentIR2026} and study how single-excitation parameter calibration can reduce the number of modes retained in explicit two-photon waveguide-QED simulations. We deliberately restrict our analysis to the Markovian regime. Although time-domain simulators already provide efficient descriptions in this setting, its well-established analytical treatment makes it a natural starting point for benchmarking the accuracy and efficiency of a frequency-domain approach. Using the monochromatic two-photon regime as a controlled benchmark, we show that a consistent reparameterization of the TLS allows the spectral window to be narrowed while preserving the consistency of the simulation, thereby reducing the dimension of the two-excitation numerical Hilbert space. The remainder of this paper is organized as follows. Section~\ref{sec:theory} introduces the theoretical framework, and Section~\ref{sec:numerical_setup} presents the numerical implementation. In Section~\ref{sec:effective_shift}, we show that effective parameter shifts persist in the two-photon regime. Section~\ref{sec:correction_prescription} then presents the proposed correction scheme, assesses its reliability for monochromatic photons, and discusses extensions to non-monochromatic inputs and connections with existing time-domain simulators.

\section{Theoretical considerations}
\label{sec:theory}

\subsection{Hamiltonian description}

\begin{figure}
    \centering
    \includegraphics[width=0.4\textwidth]{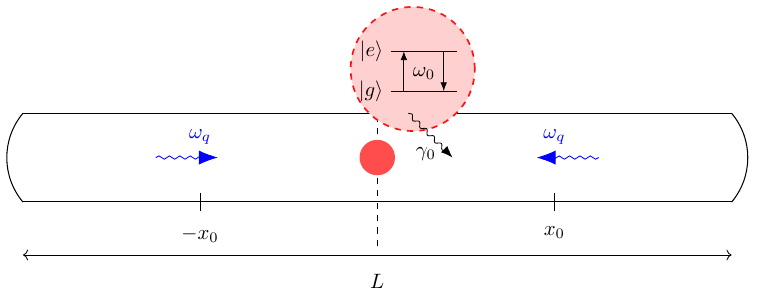}
    \caption{Schematic of the waveguide-QED system considered in this work.}
    \label{fig:waveguide_figure}
\end{figure}

We consider a waveguide QED model where, for simplicity, a single emitter modeled as a two-level system (TLS) is located at the origin $x=0$ of a one-dimensional waveguide (Fig.~\ref{fig:waveguide_figure}). The ground and excited states are denoted $\ket{g}$ and $\ket{e}$, respectively. The TLS Hamiltonian is parameterized by a transition frequency $\omega_0$ and a decay rate $\gamma_0$. The free Hamiltonian reads~\cite{royStronglyInteractingPhotons2017,pironRenormalizationTreatmentIR2026}
\begin{align}
H_0 = \int_{-\infty}^{\infty} \frac{dk}{2\pi}\, \omega_k\, a^{\dagger}(k)a(k) + \omega_0 \ket{e}\bra{e},
\end{align}
where $a^{\dagger}(k)$ and $a(k)$ create and annihilate a photon with momentum $k$ in the waveguide. Within the rotating-wave approximation (RWA), the light-matter interaction takes the form (setting $\hbar=1$)~\cite{royStronglyInteractingPhotons2017,pironRenormalizationTreatmentIR2026}
\begin{align}
V = i\sqrt{\frac{\gamma_0}{2}} \int_{-\infty}^{\infty} \frac{dk}{2\pi}\left[a(k)\ket{e}\bra{g} - \mathrm{h.c.}\right].
\end{align}
The full Hamiltonian is $H = H_0 + V$. We now discuss the form of the dispersion relation $\omega_k$.

\subsection{Treatment of the dispersion relation}

\begin{figure}
    \centering
    \includegraphics[width=0.45\textwidth]{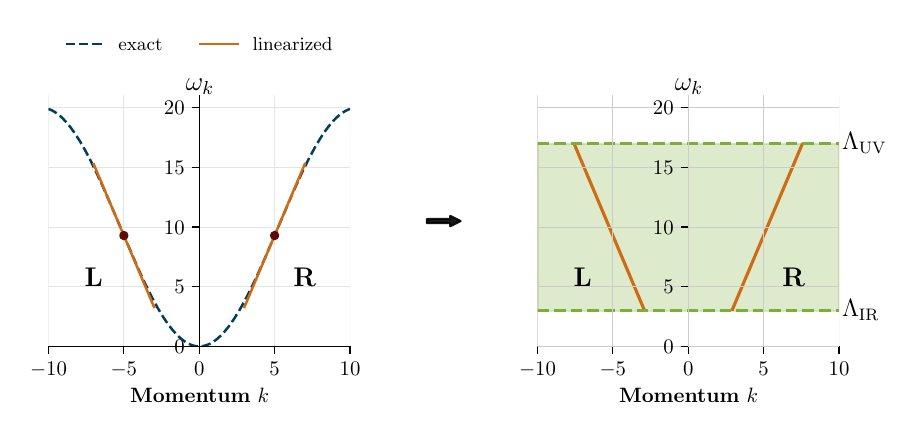}
    \caption{Linearization and truncation of the dispersion relation around the relevant set of electromagnetic modes.}
  \label{fig:dispersion_relation}    
\end{figure}

The waveguide dispersion relation $\omega_k$ can take various forms depending on the waveguide design~\cite{berndsenElectromagneticallyInducedTransparency2023,chengEngineeringPhotonicDispersion2025}, but is typically symmetric around $k=0$. It is therefore convenient to introduce a channel index $\alpha$ labeling the two propagation branches~\cite{royStronglyInteractingPhotons2017}, where $\alpha=R$ (resp.\ $\alpha=L$) corresponds to right-moving (resp.\ left-moving) modes (see left panel of Fig.~\ref{fig:dispersion_relation}). For $k>0$, one defines the single-branch operators
\begin{align}
a_R(k) = a(k), \qquad a_L(k) = a(-k).
\end{align}
Furthermore, most authors assume that the relevant physical processes occur close to a reference frequency~\cite{royStronglyInteractingPhotons2017,scullyQuantumOptics1997}. In this regime, the dispersion relation can be linearized around $\omega_{\text{ref}}$, and choosing units where the group velocity is set to unity then yields -- up to a shift of the zero-point energy -- the approximation (see left panel of Fig.~\ref{fig:dispersion_relation})
\begin{align}
    \omega_k \approx |k| \quad \text{for } k \text{ close to } \pm k_{\text{ref}}.
\end{align}
The resulting Hamiltonian can therefore be simplified and expressed solely as an integral over frequency. Going to the interaction picture, which isolates the interacting part, gives the interaction Hamiltonian~\cite{royStronglyInteractingPhotons2017,bundgaard-nielsenWaveguideQEDjlEfficientFramework2025}
\begin{align}
    V_I(t) = i \sqrt{\frac{\gamma_0}{2}} \sum_{\alpha} \int_{\substack{\vspace*{-1em }\omega \approx \omega_{\text{ref}}}} \hspace*{-1.2em }d \omega \Big[e^{-i(\omega-\omega_0)t} a_{\alpha}(\omega) \ket{e}\bra{g} - \mathrm{h.c.} \Big],
\end{align}
with $V_I(t) = e^{i H_0 t} V e^{-i H_0 t}$. The resulting model therefore defines an effective description with a spectral bandwidth centered around $\omega_{\text{ref}}$. We now turn to the standard numerical treatment of the frequency integral in the vicinity of \(\omega_{\text{ref}}\).

\subsection{Numerical representation of the photonic continuum: existing approaches}

Existing numerical approaches to waveguide QED, such as matrix product state (MPS)-based methods implemented in the library \texttt{QWaveMPS}~\cite{regidorQwaveMPSEfficientOpensource2026,arranzregidorModelingQuantumLightmatter2021}, or the recently released Julia package \texttt{WaveguideQED.jl}~\cite{bundgaard-nielsenWaveguideQEDjlEfficientFramework2025}, typically rely on a time-domain formulation of the model, whose theoretical foundations are commonly analyzed within the input-output formalism~\cite{fanInputOutputFormalismFewPhoton2010}. In these approaches, the photonic operators are expressed as~\cite{bundgaard-nielsenWaveguideQEDjlEfficientFramework2025,arranzregidorModelingQuantumLightmatter2021}
\begin{align}
    a_{\alpha}(t) = \int \frac{d\omega}{2\pi} \, e^{-i (\omega - \omega_0) t} a_{\alpha}(\omega),
    \label{eq:photon_operator_time_domain}
\end{align}
so that the interaction Hamiltonian takes the simple form
\begin{align}
    V_I(t) = i \sqrt{\frac{\gamma_0}{2}} \sum_{\alpha} 
    \Big[ a_{\alpha}(t) \ket{e}\bra{g} - a_{\alpha}^{\dagger}(t) \ket{g}\bra{e} \Big].
\end{align}
Defining the Fourier transform involved in Eq.~\ref{eq:photon_operator_time_domain} requires extending the frequency integration to the entire real axis, justified by the expectation that far-off-resonant modes have a negligible contribution~\cite{fanInputOutputFormalismFewPhoton2010, scullyQuantumOptics1997,royStronglyInteractingPhotons2017}. Consequently, the reference frequency $\omega_{\text{ref}}$ introduced above is not specified as an independent parameter of the numerical model. Instead, the convention of Eq.~\ref{eq:photon_operator_time_domain} works in a rotating frame where detunings are resolved with respect to $\omega_0$. While the retained spectral content can be characterized by the center frequency $\omega_{\text{ref}}$ and an effective bandwidth $\Lambda$, time-domain discretizations typically expose only the latter directly through the discretization time step, as we discuss in the following.

To perform numerical simulations, the resulting model must be regularized. In time domain, this is achieved by discretizing the time variable with a finite time step $\Delta t$ and restricting the simulation to a finite time window. Temporal discretization limits the range of frequencies that can be represented as dictated by the Nyquist-Shannon sampling theorem~\cite{shannonCommunicationPresenceNoise1949}. A signal sampled with time step $\Delta t$ can resolve frequencies only up to
\begin{align}
    |\omega - \omega_{\text{ref}}| \lesssim \Delta t^{-1},
\end{align}
which effectively introduces the finite spectral bandwidth
\begin{align}
    \Lambda \sim \Delta t^{-1}.
\end{align}
Accurate simulations therefore require choosing $\Delta t$ sufficiently small so that $\Lambda$ is large enough to include all relevant frequencies of the problem. However, reducing $\Delta t$ also increases the computational cost. In particular, the dimension of the numerical Hilbert space scales with the timestep, leading to a computational cost that typically grows in the two-photon sector as~\cite{bundgaard-nielsenWaveguideQEDjlEfficientFramework2025}
\begin{align}
    \dim(\mathcal{H}_{\text{num}}) \sim \Delta t^{-2} \sim \Lambda^2 .
\end{align}
These considerations highlight that computational gains can be achieved by reducing the number of photonic modes, which requires a careful analysis of the frequency representation of the model and its impact on numerical simulations. While time-domain approaches naturally constrain the photonic spectrum through $\Delta t$, the frequency-domain formulation treats the center $\omega_{\text{ref}}$ and width of the retained spectral window $\Lambda$ as independent numerical choices. In this work, we adopt a frequency-domain approach to determine how far the mode basis can be reduced and to develop a systematic framework for achieving such control.

\section{Proposed numerical setup}
\label{sec:numerical_setup}

\subsection{Resource-efficient truncation in the frequency domain}

In order to reduce the dimensionality of the Hilbert space as much as possible by keeping only the relevant frequency modes, we retain the frequency representation. The photon spectrum is discretized by considering a finite propagation length $L$ for the waveguide with periodic boundary conditions. Because we assume the approximation $\omega_k = k$ to hold within the simulation window (see right panel of Fig.~\ref{fig:dispersion_relation}), defined below, the allowed frequencies satisfy~\cite{obaFastSimulationMultiphoton2024,havukainenQuantumSimulationsOptical1999}
\begin{align}
    \omega_n = \frac{2n\pi}{L}, \quad n \in \mathbb{Z}
\label{eq:momentum_quantization}
\end{align}
Following standard procedures~\cite{obaFastSimulationMultiphoton2024,havukainenQuantumSimulationsOptical1999,pironRenormalizationTreatmentIR2026}, the Hamiltonian becomes
\begin{align} 
    \begin{split} 
        H_{\text{discretized}} &= \sum_{\alpha,n} \omega_n \, a_{\alpha n}^{\dagger} a_{\alpha n} + \omega_0 \, \ket{e}\bra{e} \\ 
        & \hspace*{3em} + i \sqrt{\frac{\gamma_0}{2L}} \sum_{\alpha,n} \Big[ a_{\alpha n} \ket{e}\bra{g} - \mathrm{h.c.} \Big],
    \end{split}
\end{align}
where the discrete operators $a_{\alpha n} = a_{\alpha}(\omega_n)/\sqrt{L}$~\cite{pironRenormalizationTreatmentIR2026} satisfy the bosonic commutation relations
\begin{align}
[a_{\alpha n},a_{\beta m}^{\dagger}] = \delta_{\alpha \beta } \delta_{mn}, \quad  [a_{\alpha n},a_{\beta m}] = [a_{\alpha n}^{\dagger},a_{\beta m}^{\dagger}] = 0.
\end{align}
We then restrict the photonic modes to a finite frequency window defined by an infrared (IR) and an ultraviolet (UV) cutoff, such that $\omega_n \in [\ir,\uv]$ (see right panel of Fig.~\ref{fig:dispersion_relation}). In practice, this restricts the allowed values of $n$ to
\begin{align}
    \frac{L}{2\pi}\ir \leq n \leq \frac{L}{2\pi}\uv.
\end{align}
The aforementioned central frequency $\omega_{\text{ref}}$ defines the spectral region of interest, while the bandwidth parameter $\Lambda$ determines the size of the simulation window,
\begin{align}
[\ir, \uv] = [\omega_{\text{ref}} - \Lambda, \, \omega_{\text{ref}} + \Lambda],
\end{align}
with
\begin{align}
    \omega_{\text{ref}} = \frac{\ir + \uv}{2}, 
    \qquad 
    \Lambda = \frac{\uv - \ir}{2}.
\end{align}
Efficient simulations require choosing $\omega_{\text{ref}}$ such that the physically relevant spectral region is captured while keeping $\Lambda$ as small as possible, as discussed in the following.

\subsection{Numerical Hilbert space}

The absence of counter-rotating terms (RWA) ensures conservation of the total excitation number~\cite{royStronglyInteractingPhotons2017}. As a consequence, an initial state with a fixed number of excitations remains confined to the corresponding invariant subspace. Focusing on two-photon scattering, we restrict the analysis to the two-excitation sector spanned by the free basis states
\begin{subequations}
    \begin{align}
        \ket{\alpha \omega_n, \beta \omega_m, g} &= a_{\alpha n}^{\dagger} a_{\beta m}^{\dagger} \ket{\emptyset_{\text{photons}}} \otimes \ket{g}, \\
        \ket{\alpha \omega_n, e} &= a_{\alpha n}^{\dagger} \ket{\emptyset_{\text{photons}}} \otimes \ket{e}.
    \end{align}
\end{subequations}
A general state is thus expanded as
\begin{align}
    \ket{\psi} =
    \sum_{\alpha n,\beta m} 
    c_{\alpha n,\beta m}
    \ket{\alpha \omega_n,\beta \omega_m,g}
    +
    \sum_{\alpha n}
    b_{\alpha n}
    \ket{\alpha \omega_n,e},
\end{align}
where it is implicitly assumed that the indices run only over frequencies contained within the finite bandwidth defined above. The corresponding numerical Hilbert space has dimension
\begin{equation}
    \dim\!\left(\mathcal{H}_{\text{num}}\right) = \underbrace{4 N_{\text{modes}}^2}_{\ket{\alpha \omega_n, \beta \omega_m, g}} + \underbrace{2 N_{\text{modes}}}_{\ket{\alpha \omega_n, e}},
    \label{eq:numerical_dimension}
\end{equation}
where
\begin{align}
    N_{\text{modes}} \sim \frac{L}{2\pi}(\uv-\ir) \propto \Lambda.
\end{align}
As expected from the previous discussion, the Hilbert space grows quadratically with the bandwidth $\Lambda$, highlighting the computational advantage of narrowing the frequency window. We therefore investigate how the simulator bandwidth can be reduced while preserving reliable physical predictions.

\subsection{Runge-Kutta propagation}

Following common approaches~\cite{havukainenQuantumSimulationsOptical1999,bundgaard-nielsenWaveguideQEDjlEfficientFramework2025,arranzregidorModelingQuantumLightmatter2021}, we work in the interaction picture, where the state $\ket{\psi(t)}$ evolves according to
\begin{align}
    i\,\partial_t \ket{\psi(t)} = V_I(t)\,\ket{\psi(t)} ,
\end{align}
with the interaction operator
\begin{align}
    V_I(t) = i \sqrt{\frac{\gamma_0}{2L}}  \sum_{\alpha, n} \Big[ e^{-i(\omega_n - \omega_0)t}\, a_{\alpha n}\,\ket{e}\bra{g} - \text{h.c.} \Big].
\end{align}
Given an initial state $\ket{\psi_0}$, the Schrödinger equation is integrated numerically using a fourth-order Runge-Kutta (RK) scheme following~\cite{havukainenQuantumSimulationsOptical1999}. This choice provides a simple and transparent implementation that remains close to the underlying physical model. The detailed construction of the algorithm is provided in Appendix~\ref{appendix:detailed_rg_scheme}. Numerical calculations were performed using \texttt{NumPy 1.26.2} with \texttt{Python 3.12.0}.

The integration over the time window $[0,T]$ yields a sequence of states $\ket{\psi_i}$ evaluated at times $t_i = i\,\Delta t$ for $i = 1, \dots, N_{\text{step}}$, where $N_{\text{step}} = T / \Delta t$. As discussed previously, a temporal discretization implicitly determines the range of frequencies that can be represented in the simulation. In the present approach, the photonic spectrum has already been restricted to the finite bandwidth $\Lambda$. Consequently, the time step only needs to be chosen such that $\Delta t = \mathcal{O}(\Lambda^{-1})$, ensuring that the relevant spectral components are properly resolved. 

The corresponding amplitudes, denoted $c_{\alpha n,\beta m, i}$ and $b_{\alpha n,i}$, provide access to the physical observables considered in this work, as discussed in the following.

\subsection{Benchmark methodology}

To quantify the accuracy of the simulator despite the frequency truncation, we benchmark it against an analytical prediction for a frequency-dependent observable extracted from the scattering dynamics. As a reference observable, we consider the coincidence probability $\mathcal{C}$, defined as the probability of detecting the two photons in different output channels~\cite{rouletTwoPhotonsAtomic2016}, for two indistinguishable photons impinging simultaneously on the scatterer (see Fig.~\ref{fig:waveguide_figure}). This quantity is commonly used in two-photon scattering experiments, as it corresponds to the anti-bunching probability associated with the Hong-Ou-Mandel (HOM) interference effect~\cite{rouletTwoPhotonsAtomic2016,hongMeasurementSubpicosecondTime1987}.

Let $(\omega_A,\gamma_A)$ denote the transition frequency and decay rate of the TLS as measured experimentally. In the monochromatic regime, where the incoming photons have a narrow frequency spread $\sigma_{\omega} \ll \gamma_A$, the TLS behaves as an effective linear scatterer~\cite{rouletTwoPhotonsAtomic2016}, and the coincidence probability has a simple analytical expression, providing a convenient benchmark for numerical simulations. This expression is~\cite{rouletTwoPhotonsAtomic2016}
\begin{equation}
    \mathcal{C}^{\text{th}}_{(\omega_A,\gamma_A)}(\omega) = 1 - 4 \mathcal{R}_{(\omega_A, \gamma_A)}(\omega)\left(1-\mathcal{R}_{(\omega_A,\gamma_A)}(\omega)\right),
    \label{eq:coincidence_theory}
\end{equation}
where $\mathcal{R}$ denotes the single-photon reflection probability
\begin{equation}
    \mathcal{R}_{(\omega_A,\gamma_A)}(\omega) =
    \frac{\gamma_A^2/4}{(\omega-\omega_A)^2 + \gamma_A^2/4}.
\end{equation}
The coincidence probability can be readily extracted from the simulation. Provided that the simulation time window $[0,T]$ is sufficiently long for the TLS to relax back to its ground state $\ket{g}$, it is obtained by evaluating the probability of finding the two photons in separate channels at the final simulation step. In the numerical model, the TLS Hamiltonian is parameterized by $(\omega_0,\gamma_0)$, leading to
\begin{align}
    \mathcal{C}^{\text{num}}_{(\omega_0, \gamma_0)}(\omega) = \mathcal{P}_{LR} + \mathcal{P}_{RL}, 
    \quad \mathcal{P}_{\alpha \beta} = \sum_{n,m} \big|c_{\alpha n, \beta m, N_{\text{step}}}\big|^2 .
\end{align} 
Reaching the monochromatic limit with high fidelity requires a strong separation of scales, which is computationally demanding in practice (see Appendix~\ref{appendix:linear_benchmark}). Hence, we quantify the simulation accuracy through a simple precision criterion: the simulated output must lie within a $5\%$ tolerance of the theoretical prediction. To evaluate this accuracy, we compute the relative error on the coincidence
\begin{align}
    \text{RE} =
    \frac{\big|\mathcal{C}^{\text{num}}_{(\omega_0, \gamma_0)}(\omega) - \mathcal{C}^{\text{th}}_{(\omega_A,\gamma_A)}(\omega)\big|}
    {\mathcal{C}^{\text{th}}_{(\omega_A,\gamma_A)}(\omega)} \; (\%),
\end{align}
where the dependence on the simulator parameters $(\omega_0,\gamma_0)$ is omitted in the following for conciseness.

\section{Finite frequency window effects}
\label{sec:effective_shift}

A finite frequency window was shown to shift the effective parameters of single-photon scattering simulations~\cite{pironRenormalizationTreatmentIR2026}. In this section, we demonstrate through benchmark simulations that similar effects persist in the present setup.

\subsection{Input state initialization}

To approach the monochromatic limit while keeping the computational cost reasonable (see Appendix~\ref{appendix:linear_benchmark}), we fix the parameters as reported in Table~\ref{tab:simulation_parameters} and describe below how the photon state is initialized. Time evolution is performed using the RK scheme up to $T = L/2 = 25$ with $N_{\text{step}} = 2500$ (time step $\Delta t = 0.01$), ensuring numerical convergence over a wide range of bandwidths $\Lambda$.

At the initial time, the system contains two photons while the TLS is prepared in its ground state. The photons are indistinguishable and initialized as Gaussian wave packets centered at frequency $\omega_q$ with standard deviation $\sigma_{\omega}$ in frequency space. They are injected into distinct propagation channels. In position space, the packets are centered at $x_0$, which can equivalently be interpreted as a propagation delay as the group velocity is set to unity. The initial state thus reads
\begin{align}
    \ket{\psi_0} =
    \sum_{\alpha n, \beta m}
    c_{\alpha n, \beta m, 0}\, \ket{\alpha \omega_n, \beta \omega_m, g} ,
\end{align}
with coefficients~\cite{obaFastSimulationMultiphoton2024}
\begin{align} 
    \begin{split} 
        c_{\alpha n, \beta m, 0} \sim &
        \Bigg\{\delta_{\alpha L}\,
        \exp\!\left[ -\frac{(\omega_n-\omega_q)^2}{4\sigma_{\omega}^2} - i \omega_n x_0 \right] \\
        & \hspace*{-5em}\times \delta_{\beta R}\,
        \exp\!\left[ -\frac{(\omega_m-\omega_q)^2}{4\sigma_{\omega}^2} - i \omega_m x_0 \right]
        \Bigg\}
        + \Big\{ (\alpha n) \leftrightarrow (\beta m) \Big\},
    \end{split} 
\end{align}
up to a normalization factor, which is computed numerically as a function of $\ir$ and $\uv$ to ensure proper normalization of the wave packets.
\begin{table}[h]
\centering
\begin{tabular}{|l l|}
\hline
\multicolumn{2}{|c|}{\textbf{Waveguide parameters}} \\
\hline
Atomic frequency & $\omega_A = 10\pi$ \\
Atomic decay rate & $\gamma_A = 5\pi$ \\
Length & $L = 50$ \\
\hline
\multicolumn{2}{|c|}{\textbf{Photon parameters}} \\
\hline
Incoming frequency & $\omega_q$ (variable) \\
Frequency spread & $\sigma_{\omega} = 10^{-2}\gamma_A = 0.05\pi$ \\
Initial position & $x_0 = -L/4 = -12.5$ \\
\hline
\end{tabular}
\vspace*{1em}
\caption{Parameters used to approach the monochromatic limit.}
\label{tab:simulation_parameters}
\end{table}

\begin{figure*}[!t]
    \centering
    \includegraphics[width=\textwidth]{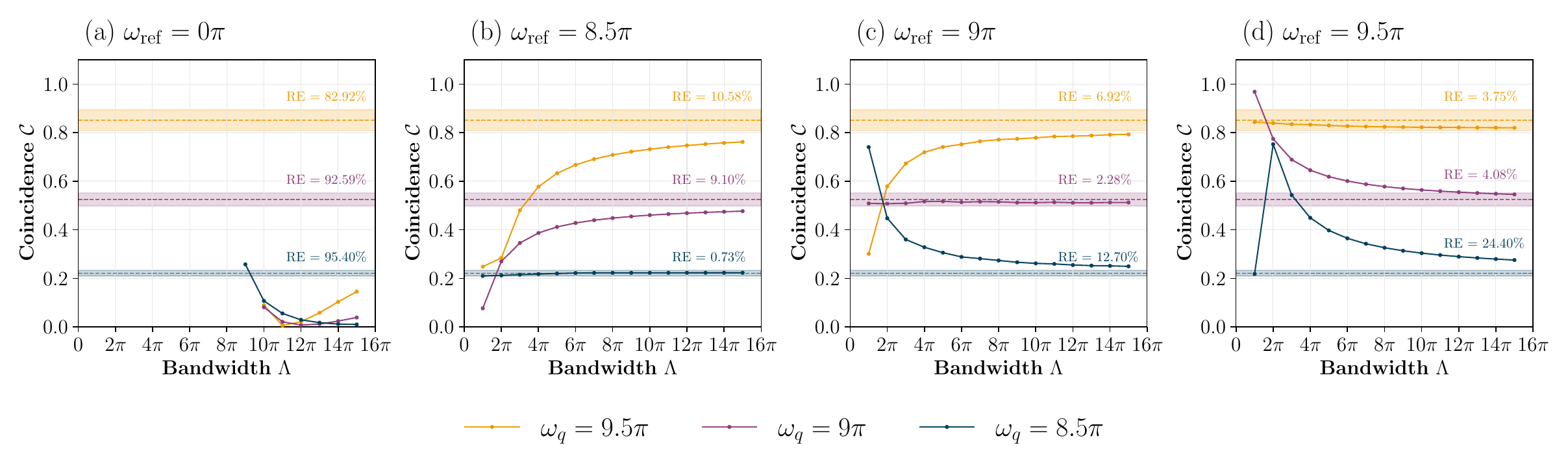}
    \vspace*{-2em}
    \caption{Coincidence against the simulation bandwidth for different incoming photon frequencies. The frequency windows are centered at $\omega_{\text{ref}} = 0\pi$ (a), $8.5\pi$ (b), $9\pi$ (c), and $9.5\pi$ (d)}
    \label{fig:coincidence_vs_bandwidth_baseline}
\end{figure*}
\begin{figure*}[t]
    \centering
    \includegraphics[width=\textwidth]{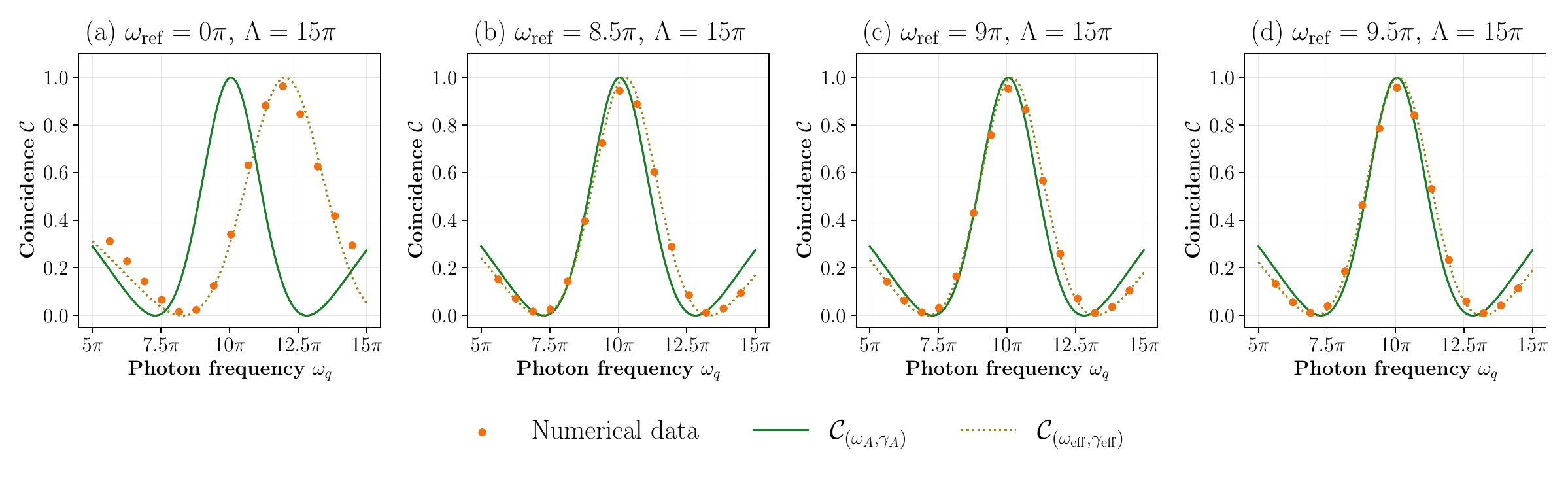}\vspace*{-2em}
    \caption{Coincidence (orange points) against incoming photon frequency for windows centered at $\omega_{\text{ref}} = 0\pi$ (a), $8.5\pi$ (b), $9\pi$ (c), and $9.5\pi$ (d), bandwidth $\Lambda = 15\pi$ across all cases. The expected theoretical curve is shown in green, and the effective fit with a dashed line.}
    \label{fig:coincidence_vs_frequency}
\end{figure*}

\subsection{Bandwidth dependence of the baseline simulator}

We first check the reliability of the simulator when the TLS parameters are set to their experimentally measured values, i.e., $(\omega_0, \gamma_0) = (\omega_A, \gamma_A)$. To probe the impact of bandwidth truncation across the photon spectrum, we consider three photon frequencies slightly detuned from the atomic transition, $\omega_q = 9.5\pi,\, 9\pi,$ and $8.5\pi$. For several choices of the central frequency $\omega_{\mathrm{ref}} \in \{0\pi,\, 8.5\pi,\, 9\pi,\, 9.5\pi\}$, the simulation bandwidth $\Lambda$ is progressively increased (see Appendix~\ref{appendix:negative_freq} for a discussion of the inclusion of negative frequencies when $\Lambda > \omega_{\mathrm{ref}}$). The choice $\omega_{\mathrm{ref}} = 0\pi$ is included as a deliberately miscentered reference case, far from the relevant energy scales, and is expected to require a larger bandwidth. The other values are matched to the probe frequencies and are therefore expected to yield improved accuracy.

The results are reported in Fig.~\ref{fig:coincidence_vs_bandwidth_baseline}, restricting the analysis to configurations where $\omega_q$ lies within the simulation window, as no meaningful data can otherwise be extracted. The plot reveals that increasing the bandwidth does not systematically yield convergence of the coincidence probability towards a plateau, although $\Lambda = 15\pi$ already covers all relevant frequencies of the underlying process. In particular, the miscentered window with $\omega_{\mathrm{ref}} = 0\pi$ exhibits no convergence within the explored bandwidth range: even at $\Lambda = 15\pi$, the relative error remains very large (exceeding $80\%$ in Fig.~\ref{fig:coincidence_vs_bandwidth_baseline}(a)). Although convergence improves for $\omega_{\mathrm{ref}} = \omega_q$ (Fig.~\ref{fig:coincidence_vs_bandwidth_baseline}(b-d)), this remains frequency-local and fails to capture the response at other $\omega_q$.

These observations reveal a fundamental limitation of the naive truncation strategy. In principle, a sufficiently large bandwidth should restore the correct scattering response independently of the choice of $\omega_{\mathrm{ref}}$. However, this convergence is not observed: the results remain strongly dependent on $\omega_{\mathrm{ref}}$, and miscentered choices lead to large residual errors even at large bandwidth.

\subsection{Emergence of effective parameters}

To further investigate the origin of this non-convergence issue, we perform an additional series of scattering simulations for the same central frequencies $\omega_{\text{ref}} \in \{0\pi, 8.5\pi, 9\pi, 9.5\pi\}$ at the largest bandwidth previously considered, $\Lambda = 15\pi$. We vary the incoming photon frequency $\omega_q$ over a broad range to probe the functional dependence of the coincidence. The Hamiltonian parameters used in the simulation remain fixed to $(\omega_0, \gamma_0) = (\omega_A, \gamma_A)$.

Interestingly, the data in Fig.~\ref{fig:coincidence_vs_frequency} do not follow the expected curve $\mathcal{C}_{(\omega_A,\gamma_A)}(\omega)$, although the simulator is initialized with $(\omega_0,\gamma_0)=(\omega_A,\gamma_A)$. Instead, they are well described by the same functional form evaluated at a set of \emph{effective} parameters $(\omega_{\text{eff}},\gamma_{\text{eff}})$. Notably, for each window, the effective curve may intersect the theoretical one, implying that the apparent agreement observed previously was in fact accidental.

This suggests that simulations at reduced bandwidth should still be feasible in this example ($\Lambda < 15\pi$) by appropriately adjusting the simulator parameters. More precisely, consistently adjusting $(\omega_0, \gamma_0)$ to enforce the condition
\begin{align}
    \mathcal{C}_{(\omega_A,\gamma_A)} = \mathcal{C}_{(\omega_{\text{eff}},\gamma_{\text{eff}})},
\end{align}
should enable reduced-bandwidth simulations. However, in Fig.~\ref{fig:coincidence_vs_frequency}, the parameters $(\omega_{\text{eff}},\gamma_{\text{eff}})$ are obtained via a direct least-squares fit of the numerical data using \texttt{scipy.optimize.curve\_fit} (v1.11.3), purely for illustrative purposes. A practical use of the simulator would require a more systematic procedure to determine these parameters. We now turn to a practical strategy to implement this idea, building on preliminary work~\cite{pironRenormalizationTreatmentIR2026}.

\section{Calibration prescription for accurate reduced-bandwidth simulations}
\label{sec:correction_prescription}

\subsection{Single-excitation bandwidth corrections}

The theoretical analysis of~\cite{pironRenormalizationTreatmentIR2026} shows how to simulate a single photon scattering experiment on a TLS with physical parameters $(\omega_A, \gamma_A)$ by appropriately tuning the numerical parameters $(\omega_0,\gamma_0)$. This procedure relies on relating them to the shift of the poles of the atomic Green function induced by the finite bandwidth inherent to numerical simulations. 

Assuming that $\omega_0$ is sufficiently detuned from the IR and UV cutoffs $\ir$ and $\uv$, the TLS kernel can be treated as approximately Markovian, as discussed in~\cite{pironRenormalizationTreatmentIR2026}. In this regime, the kernel admits a local expansion, which in turn enables a self-consistent determination of the pole $\lambda$ of the atomic Green's function~\cite{pironRenormalizationTreatmentIR2026}:
\begin{align}
    \lambda = \sum_{k \geq 0} \alpha_k \lambda^k,
    \qquad 
    \text{with }
    \lambda = i(\omega_0 - \omega_A) - \frac{\gamma_A}{2},
    \label{eq:self_consistent_eq_series}
\end{align}
with the expansion coefficients given by
\begin{align}
    \begin{split}
        \alpha_0 &= -\dfrac{\gamma_0}{2} + i \dfrac{\gamma_0}{2\pi} \log\!\left(\dfrac{\uv - \omega_0}{\omega_0 - \ir}\right), \\
        \alpha_{k \geq 1} &= i^{\,k-1} \dfrac{\gamma_0}{2 k \pi} \left[ \dfrac{1}{(\uv - \omega_0)^k} + \dfrac{(-1)^{k-1}}{(\omega_0 - \ir)^k} \right].
    \end{split}
    \label{eq:alpha_coef}
\end{align}
This equation is of practical interest, as it links numerical parameters $(\omega_0, \gamma_0)$ to physical ones $(\omega_A, \gamma_A)$. The pair $(\omega_0, \gamma_0)$ can then be obtained by truncating the series to order $k \leq K$ and numerically finding a suitable root within the interval $[\ir, \uv]$. We denote the resulting parameterization by $(\omega_A^{(K)}, \gamma_A^{(K)})$. We now assess whether this parameterization remains valid in the two-photon sector.

\subsection{Minimal truncation order for two-photon simulations}

\begin{figure}
    \centering
    \includegraphics[width=0.47\textwidth]{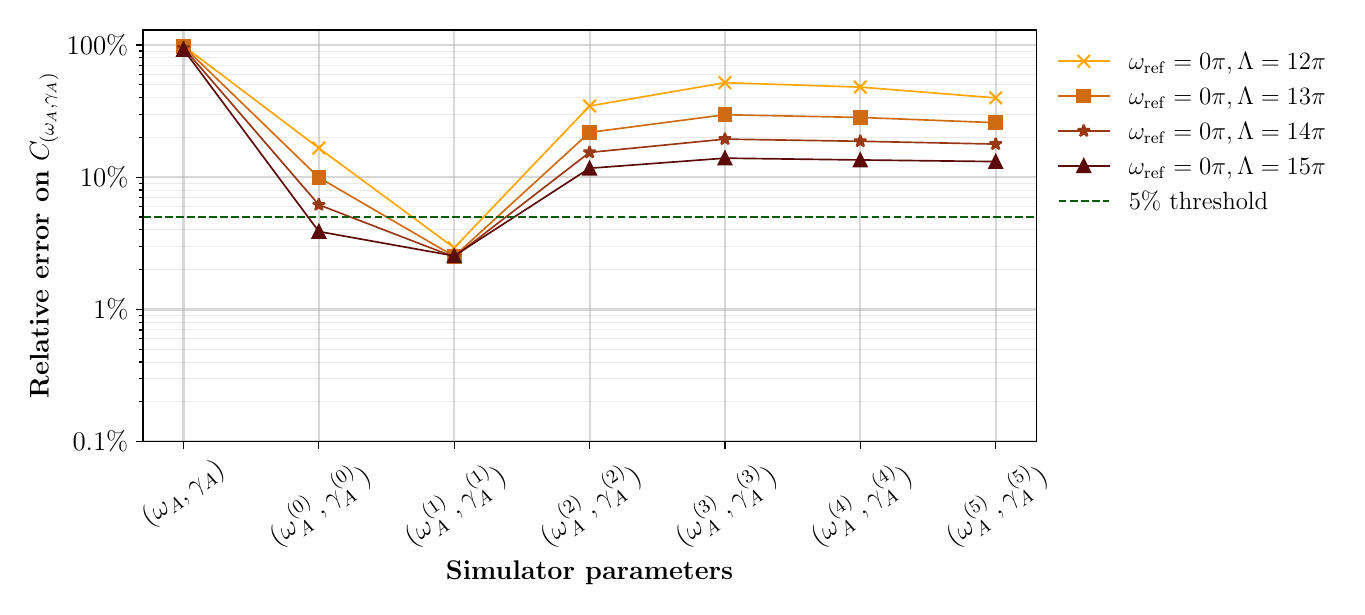}
    \caption{Relative error on the coincidence at $\omega_q = 9\pi$ for different parameter sets $(\omega_0,\gamma_0)$. Each curve corresponds to a different frequency window.}
    \label{fig:coincidence_vs_N}
\end{figure}

\begin{figure*}[t]
    \centering
    \includegraphics[width=\textwidth]{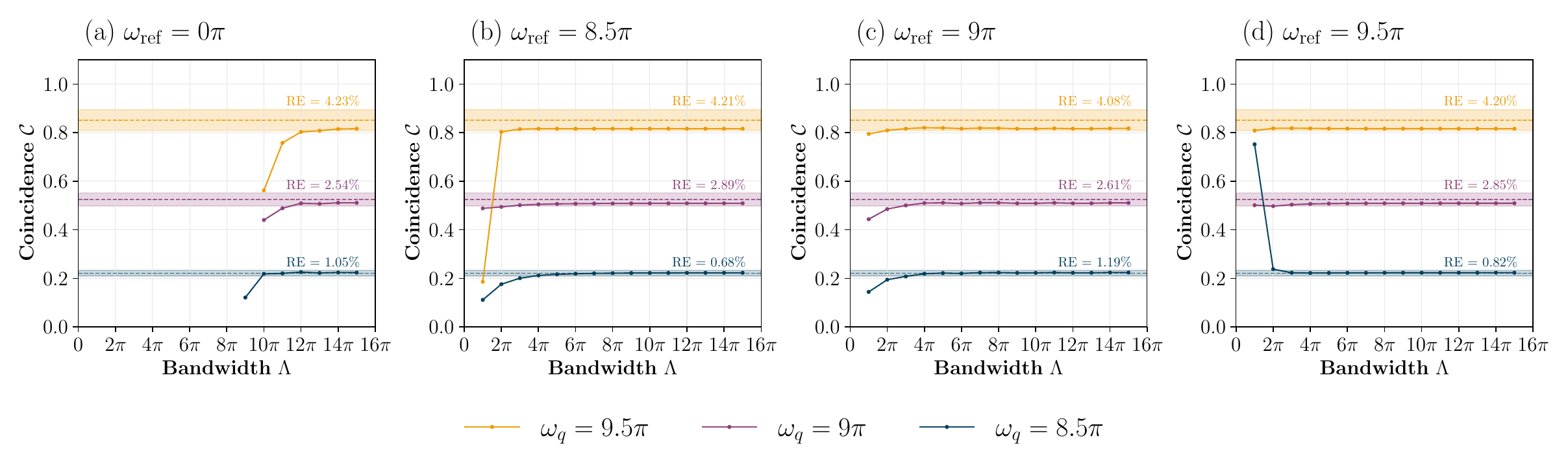}
    \vspace*{-2em}
    \caption{Convergence of the coincidence restored after calibration, as a function of the simulation bandwidth for different incoming photon frequencies. The frequency windows are centered at $\omega_{\text{ref}} = 0\pi$ (a), $8.5\pi$ (b), $9\pi$ (c), and $9.5\pi$ (d)}
    \label{fig:coincidence_vs_bandwidth_corrected}
\end{figure*}

\begin{figure*}[t]
    \centering
    \includegraphics[width=\textwidth]{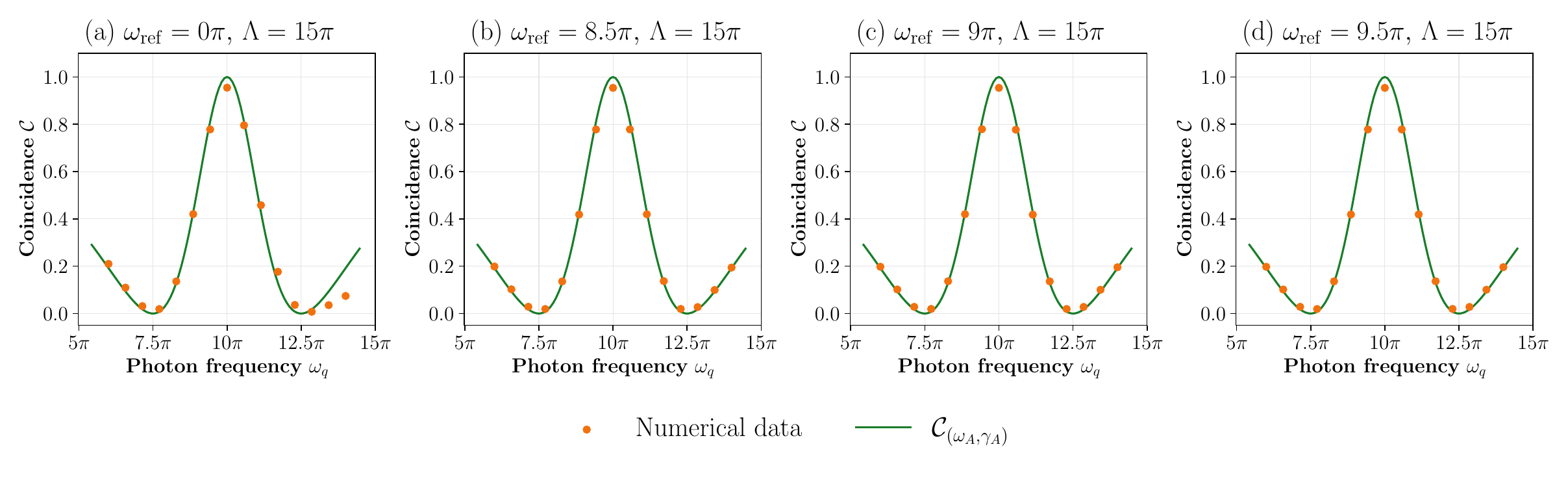}\vspace*{-2em}
    \caption{Consistency restored between simulated data (orange markers) and the theoretical coincidence as a function of the incoming photon frequency. Results are shown for frequency windows centered at $\omega_{\text{ref}} = 0\pi$ (a), $8.5\pi$ (b), $9\pi$ (c), and $9.5\pi$ (d), bandwidth $\Lambda = 15\pi$ across all cases.}
    \label{fig:coincidence_vs_frequency_corrected}
\end{figure*}

We consider a representative photon frequency $\omega_q = 9\pi$, together with the worst-case configuration identified in the previous benchmark, $\omega_{\text{ref}} = 0\pi$. To probe whether the previously considered bandwidths already allow for accurate simulations after corrections, we examine $\Lambda = 12\pi, 13\pi, 14\pi, 15\pi$, thereby covering the upper range of values explored earlier.

We evaluate the coincidence through simulations using different parameterizations of the TLS Hamiltonian, while all other settings, including the RK propagation scheme, are kept identical to those used previously (see Table~\ref{tab:simulation_parameters}). We compare the uncorrected baseline $(\omega_0,\gamma_0)=(\omega_A,\gamma_A)$ with the corrected parameters $(\omega_A^{(K)},\gamma_A^{(K)})$ obtained from the proposed scheme, with $K=0,\dots,5$.

The results are shown in Fig.~\ref{fig:coincidence_vs_N}, where the $5\%$ relative error threshold is indicated. As expected, the uncorrected baseline exhibits large errors, reaching values close to $100\%$, consistent with previous observations. In contrast, truncation at $K=1$ brings all considered bandwidths below the target accuracy, confirming that the bandwidth is sufficient and that the dominant error previously observed actually arose from miscalibration of the simulation.

For $K>1$, the accuracy deteriorates again. This behavior can be explained by the limited separation between the UV cutoff and the atomic frequency (here: $\omega_A = 10\pi, \omega_{\text{ref}} = 0\pi, \Lambda = 12\pi,\dots,15\pi$), which induces non-Markovian corrections to the TLS numerical kernel. In this regime, the local expansion underlying Eq.~\ref{eq:self_consistent_eq_series} becomes less reliable, and higher-order truncations no longer improve the approximation. Overall, these results support the prescription $K=1$.

\subsection{Restoring accuracy through parameter calibration}

\begin{algorithm}
\caption{Calibration procedure}
\label{alg:param_proposal}
\begin{algorithmic}[1]

\Require Physical parameters $(\omega_A, \gamma_A)$
\Ensure TLS simulator parameters $(\omega_0, \gamma_0)$

\State Fix $\omega_{\text{ref}}$, $\Lambda$ 
\State Compute $\omega_A^{(1)}$ by solving numerically
$$
\omega_A = \omega_A^{(1)} - \frac{\gamma_A}{2\pi} \log \left( \frac{\Lambda + (\omega_{\text{ref}} - \omega_A^{(1)})}{\Lambda - (\omega_{\text{ref}} - \omega_A^{(1)})} \right)
$$
\State Deduce $\gamma_A^{(1)}$ according to
$$
    \gamma_A^{(1)} = \gamma_A \left[ 1 + \frac{\gamma_A}{\pi} \frac{\Lambda}{\Lambda^2 - \left(\omega_A^{(1)} - \omega_{\text{ref}}\right)^2} \right]^{-1}
$$

\Return $(\omega_0, \gamma_0) = (\omega_A^{(1)}, \gamma_A^{(1)})$

\end{algorithmic}
\end{algorithm}

Given a frequency window, we prescribe solving the self-consistent equation at truncation order $K=1$ to parameterize a reduced-bandwidth simulation. Using the expressions for $\alpha_0$ and $\alpha_1$ (Eq.~\ref{eq:alpha_coef}), a parameter selection scheme can be derived and is summarized in Algorithm~\ref{alg:param_proposal}.

To assess this proposal, we revisit the previous benchmark simulations using the corrected parameterization. Following the same methodology as before, we consider photon frequencies $\omega_q = 9.5\pi,\, 9\pi,$ and $8.5\pi$, and study the coincidence as a function of the bandwidth for central frequencies $\omega_{\text{ref}} \in \{0\pi,\, 8.5\pi,\, 9\pi,\, 9.5\pi\}$. The simulator is now parameterized with $(\omega_0, \gamma_0) = (\omega_A^{(1)}, \gamma_A^{(1)})$. 

The results shown in Fig.~\ref{fig:coincidence_vs_bandwidth_corrected} exhibit the expected consistency: all curves converge to a common plateau value at sufficiently large bandwidth, which can be interpreted as the simulator prediction. However, the bandwidth required to reach this regime strongly depends on the choice of $\omega_{\mathrm{ref}}$. In the extreme case $\omega_{\mathrm{ref}} = 0\pi$, convergence is only achieved for $\Lambda \gtrsim 12\pi$, whereas for central frequencies matched to the relevant energy scales (Fig.~\ref{fig:coincidence_vs_bandwidth_corrected}(b,c,d)), a much smaller bandwidth $\Lambda \sim 2\pi$--$4\pi$ is sufficient for all curves to lie within the $5\%$ confidence region. This demonstrates that, while the choice of $\omega_{\mathrm{ref}}$ does not alter the asymptotic prediction of the simulator, it has a major impact on the computational resources required to reach convergence. In the present case, adjusting $\omega_{\text{ref}}$ allows reducing the bandwidth by a factor of 4, which translates into a 16-fold reduction of the effective Hilbert space dimension, owing to its quadratic scaling (see Eq.~\ref{eq:numerical_dimension}).

We also reproduced the second benchmark using the numerical TLS parameters $(\omega_0, \gamma_0) = (\omega_A^{(1)}, \gamma_A^{(1)})$, where the coincidence is evaluated as a function of the photon frequency $\omega_q$ for central frequencies $\omega_{\text{ref}} \in \{0\pi, 8.5\pi, 9\pi, 9.5\pi\}$, with the bandwidth fixed to $\Lambda = 15\pi$. As shown in Fig.~\ref{fig:coincidence_vs_frequency_corrected}, all four panels now match the theoretical curve, as expected since $\Lambda = 15\pi$ captures all relevant energy scales. A slight deviation is observed at high frequencies in subpanel~(a), due to the proximity to the UV cutoff. Overall, this confirms that the proposed prescription enables tuning the spectral window to improve computational efficiency while preserving the physical consistency of the simulation.

\subsection{Limitations and discussion}

Two questions naturally arise regarding the proposed calibration procedure (Alg.~\ref{alg:param_proposal}). First, does it remain relevant for non-monochromatic photons? Second, why do time-domain simulations targeting the same regime as the present work not require such renormalization-aware calibration? We discuss these two points below.

\begin{figure*}[t]
    \centering
    \includegraphics[width=\textwidth]{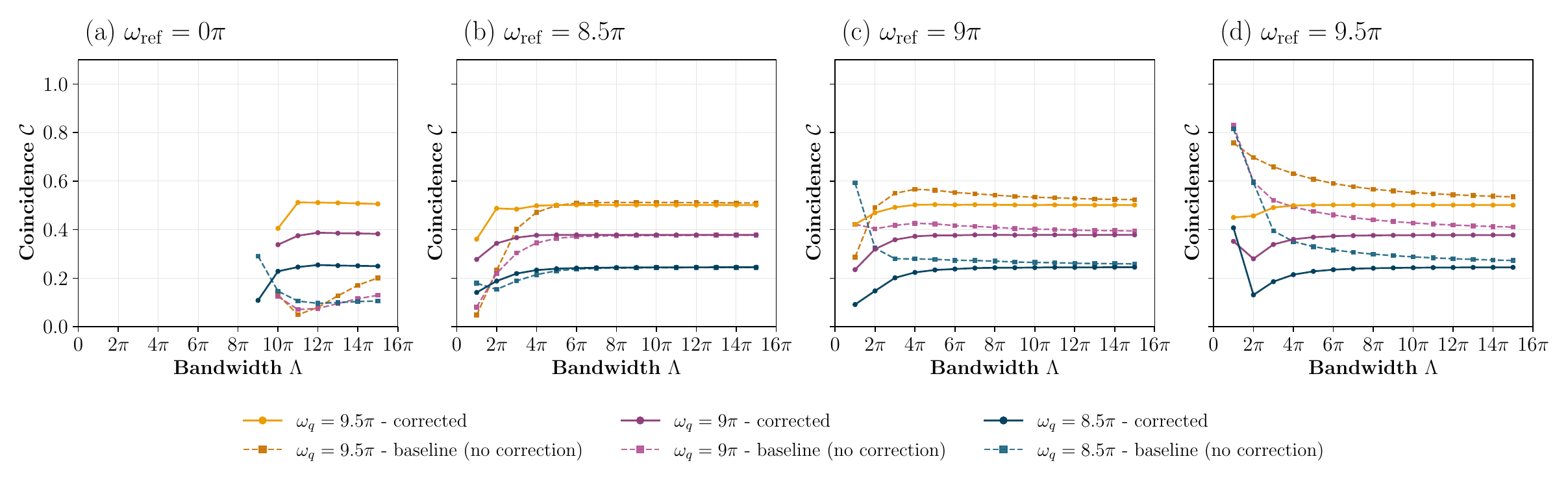}
    \vspace*{-2em}
    \caption{Convergence of the coincidence probability for non-monochromatic photons as a function of the simulation bandwidth. Dashed lines show uncalibrated simulations, while solid lines show calibrated results. The frequency windows are centered at $\omega_{\mathrm{ref}}=0\pi$ (a), $8.5\pi$ (b), $9\pi$ (c), and $9.5\pi$ (d).}
    \label{fig:coincidence_vs_bandwidth_non_monochr_consistency}
\end{figure*}

\begin{figure}
    \centering
    \includegraphics[width=0.48\textwidth]{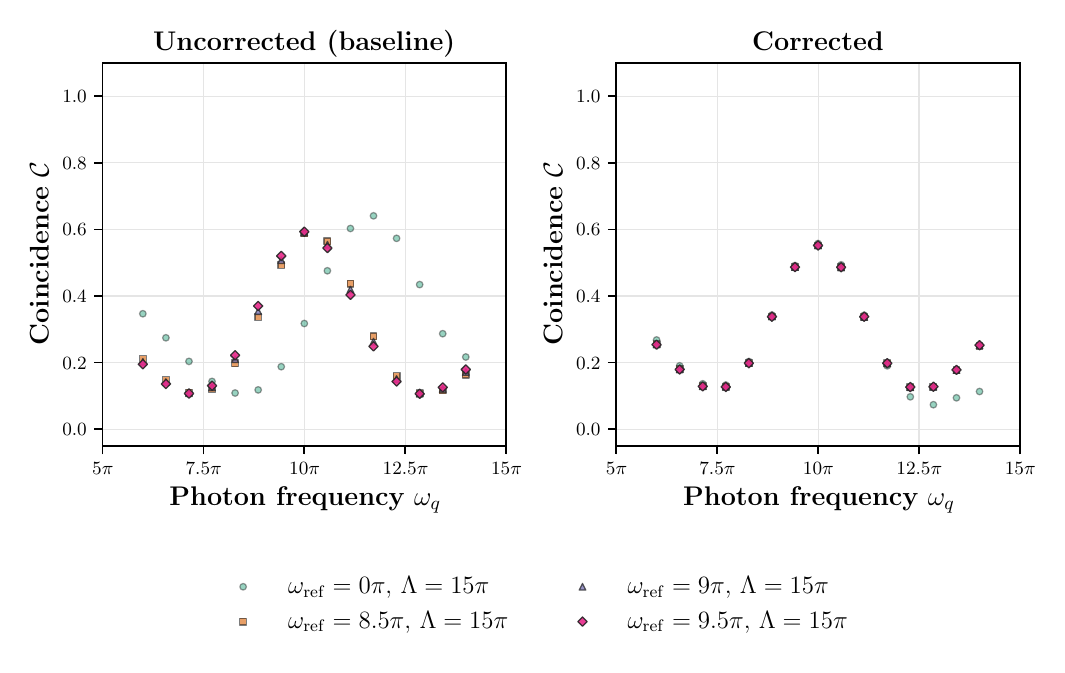}
    \caption{Coincidence probability as a function of the incoming photon frequency for the four simulation windows used in Figs.~\ref{fig:coincidence_vs_frequency}, \ref{fig:coincidence_vs_frequency_corrected}, without calibration (left) and with calibration (right).}
  \label{fig:coincidence_vs_frequency_non_monochr_comparison}    
\end{figure}

\subsubsection{Non-monochromatic wavepackets}
The previous experiments demonstrating the relevance of Alg.~\ref{alg:param_proposal} were performed using monochromatic photons, where the TLS linear response provides a controlled analytical reference. While this is sufficient to demonstrate that finite-bandwidth renormalization persists in two-photon simulations and that calibration restores the expected observables in this regime, it does not ensure that all finite-bandwidth effects are corrected for arbitrary wave packets. This limitation can be understood from the structure of the two-photon equations of motion, which involve both a single-excitation kernel $K^{(1)}$ and a genuinely multi-photon contribution $K^{(2)}$ (see Appendix~\ref{appendix:kernels}). The proposed calibration corrects the linear component embedded in the two-photon response, but does not explicitly renormalize $K^{(2)}$. A complete analysis of non-monochromatic wave packets would therefore require an analytical baseline for the full two-photon response (see, e.g.,~\cite{shenStronglyCorrelatedTwoPhoton2007}), and is left for future work. Nevertheless, consistency checks across frequency windows can still assess whether the calibration remains beneficial beyond the monochromatic limit.

Specifically, we repeated the two benchmark tests with indistinguishable Gaussian photons using $\sigma_{\omega}=0.5\pi$, ten times larger than in Table~\ref{tab:simulation_parameters}, while keeping all other parameters unchanged. Figure~\ref{fig:coincidence_vs_bandwidth_non_monochr_consistency} shows the coincidence probability as a function of the bandwidth for different central frequencies. In the uncorrected simulations, the behavior is less unstable than in the monochromatic benchmark, in particular $\omega_{\mathrm{ref}}=8.5\pi$ already displays reasonable convergence. Nevertheless, the limiting values remain visibly dependent on the chosen frequency window, for example $\omega_{\mathrm{ref}}=0\pi$ still fails to converge towards a reliable prediction when the simulation is not corrected. In contrast, after applying the calibration prescription, the curves converge toward consistent values across all central frequencies, and typically do so at smaller bandwidths.

Figure~\ref{fig:coincidence_vs_frequency_non_monochr_comparison} shows the spectral profile of the coincidence, where the bandwidth is fixed to $\Lambda=15\pi$ and the incoming photon frequency is varied. Without calibration, different choices of $\omega_{\mathrm{ref}}$ lead to different spectral profiles of the coincidence probability, indicating that the simulation still retains an artificial dependence on the numerical window. After calibration, the profiles collapse onto a common response, up to a small deviation for the window centered at $\omega_{\mathrm{ref}}=0\pi$, where the highest-frequency points lie close to the UV cutoff. These observations suggest that the single-excitation calibration remains useful beyond the monochromatic limit: although it does not provide a complete renormalization of the full two-photon dynamics, it substantially reduces the dependence on the finite frequency window.

\subsubsection{Connection with time-domain approaches}

\begin{figure}
    \centering
    \includegraphics[width=0.48\textwidth]{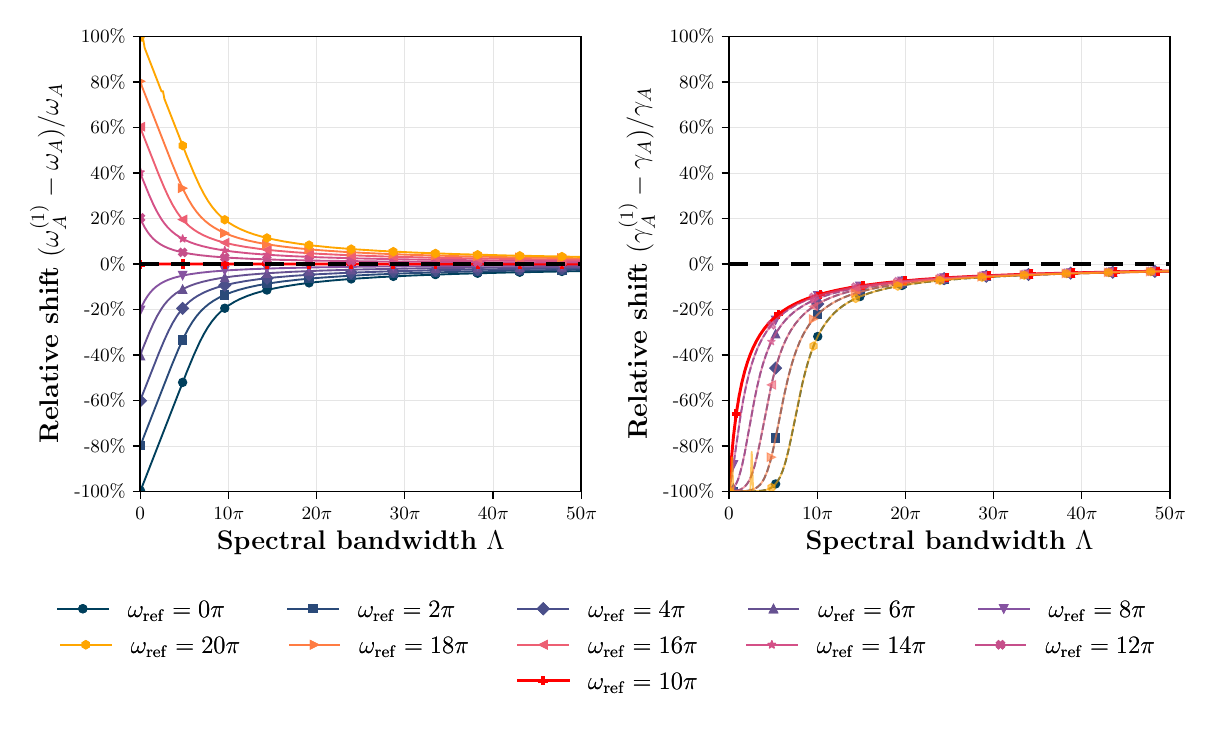}
    \caption{Relative shifts of $\omega_A^{(1)}$ and $\gamma_A^{(1)}$ with respect to $\omega_A$ and $\gamma_A$ as functions of the bandwidth, for different central frequencies. The left panel shows frequency shifts, while the right panel shows decay-rate shifts.}
  \label{fig:correction_deviation}    
\end{figure}

The finite-bandwidth effects identified in this work should not be specific to the frequency-domain representation. Time-domain simulators implicitly impose a finite cutoff through the time-discretization step~\cite{shannonCommunicationPresenceNoise1949}, $\Lambda \sim \Delta t^{-1}$, yet existing time-domain approaches to waveguide QED rarely discuss them~\cite{bundgaard-nielsenWaveguideQEDjlEfficientFramework2025,arranzregidorModelingQuantumLightmatter2021,regidorQwaveMPSEfficientOpensource2026,khanTensorNetworkApproach2025}, raising the question of why this is the case. To gain insight, we analyze the corrected parameters $\omega_A^{(1)}$ and $\gamma_A^{(1)}$ as functions of $\Lambda$ for fixed physical values $\omega_A = 10\pi$ and $\gamma_A = 5\pi$. As shown in Fig.~\ref{fig:correction_deviation}, both corrections vanish in the large-bandwidth limit,
\begin{align}
    \omega_A^{(1)} \to \omega_A, \quad \gamma_A^{(1)} \to \gamma_A \quad \text{as } \Lambda \to \infty,
\end{align}
which likely explains why recalibration is usually avoided in time-domain simulations: choosing a sufficiently small $\Delta t$ effectively corresponds to a large bandwidth $\Lambda$, in which case the corrected parameters $(\omega_0, \gamma_0) = (\omega_A^{(1)}, \gamma_A^{(1)})$ coincide with the uncorrected baseline $(\omega_0, \gamma_0) = (\omega_A, \gamma_A)$. However, decreasing $\Delta t$ increases the number of resolved temporal degrees of freedom, and therefore the numerical cost. Furthermore, the minimal bandwidth required to suppress these corrections increases with the detuning between $\omega_{\text{ref}}$ and $\omega_A$, suggesting that the implicit choice of reference frequency in time-domain approaches impacts numerical efficiency. Overall, this suggests that an explicit finite-bandwidth analysis in the time domain could help identify when larger time steps remain accurate, but such a validation is left for future work.

\section{Conclusion}

\subsection{Summary of contributions}
In this work, we have:
\begin{enumerate}[label=(\roman*)]
    \item Introduced a reduced-bandwidth frequency-domain formulation of waveguide-QED simulations in which both the bandwidth and the central frequency of the numerical window are explicit control parameters. We have shown that hard spectral truncation then induces effective shifts of the TLS parameters, and we provided a practical calibration that compensates for these effects.
    \item Demonstrated on a controlled monochromatic two-photon benchmark that this prescription can significantly reduce the required bandwidth while preserving the scattering observables. In the example considered here, a fourfold bandwidth reduction yields a sixteenfold reduction of the effective Hilbert space dimension.
\end{enumerate}

\subsection{Future work}
Several directions for future work can be identified. First, it would be valuable to further investigate the validity of the proposed calibration strategy beyond the monochromatic limit, using available analytical results in the nonlinear regime as references~\cite{schneiderGreensFunctionFormalism2016a,shenStronglyCorrelatedTwoPhoton2007}. Second, a similar analysis could be carried out in the time domain to reduce the number of time bins in existing simulators~\cite{bundgaard-nielsenWaveguideQEDjlEfficientFramework2025,regidorQwaveMPSEfficientOpensource2026}, thereby improving their efficiency in the same spirit as the frequency-domain approach developed here.

\section*{Code availability}

The code used to generate the numerical results presented in this work is publicly available at \url{https://github.com/rpiron/calibration_twophoton_waveguideQED}.

\section*{Acknowledgment}

We thank Victor M. Bastidas for insightful remarks regarding the connection between UV cutoffs and time-discretization effects. GPT-5.3 was used to assist in generating and formatting the plots in Figs. [3-9]. The authors provided data and plotting instructions. The final figures were reviewed and verified by the authors. This work was supported by JST Moonshot R\&D Program Grant Numbers JPMJMS226C and JPMJMS256K.

\bibliographystyle{IEEEtran}
%\nocite{*}
\bibliography{library}

\appendices

\section{On the numerical cost of the monochromatic limit}
\label{appendix:linear_benchmark}

The regime $\sigma_{\omega} \ll \gamma_A$ provides an analytically controlled benchmark (linear response of the TLS) for assessing the simulator's accuracy, but approaching this limit numerically involves a nontrivial tradeoff. The wave packet must remain well localized in time domain, $\sigma_t \ll L$, while for a Gaussian wave packet one has $\sigma_{\omega} \sigma_t = 1/2$ (with $\hbar = 1$), which implies $\sigma_{\omega} \gg 1/L$. Consistency therefore requires operating in the window
\begin{align}
    \frac{1}{L} \ll \sigma_{\omega} \ll \gamma_A.
\end{align}
Achieving this separation of scales necessitates increasing either $\gamma_A$ or $L$, both of which significantly raise the numerical cost. Increasing $L$ directly increases the number of modes, as
\begin{align}
    N_{\text{modes}} \sim \frac{L}{2\pi}(\uv-\ir),
\end{align}
while increasing $\gamma_A$ necessitates enlarging the bandwidth to include frequencies down to $\omega_A - \gamma_A/2$ (see Eq.~\ref{eq:coincidence_theory}). Both strategies therefore increase the required mode basis, making the monochromatic limit computationally demanding despite its conceptual simplicity.

\section{Inclusion of negative frequencies in the numerical model}
\label{appendix:negative_freq}

Although the frequency-domain photonic Hamiltonian is defined over positive frequencies only, time-domain approaches rely on field operators defined through Fourier transforms of the form~\cite{bundgaard-nielsenWaveguideQEDjlEfficientFramework2025,regidorQwaveMPSEfficientOpensource2026,arranzregidorModelingQuantumLightmatter2021}
\begin{equation}
a(t) = \int_{-\infty}^{+\infty} d\omega \, a(\omega) e^{-i \omega t},
\end{equation}
which formally extend the frequency domain to negative values, commonly justified by the fact that highly detuned frequency components have a negligible contribution to the dynamics~\cite{scullyQuantumOptics1997}. In the present simulations, we allow for negative frequencies in order to remain consistent with standard numerical approaches to waveguide-QED.

\section{Runge-Kutta vectorization}
\label{appendix:detailed_rg_scheme}

We detail here the derivation of the equations used to propagate the system state within the fourth-order RK scheme~\cite{havukainenQuantumSimulationsOptical1999}. The interaction time window is discretized as $t_i = i\,\Delta t$, and the state is updated according to
\begin{align} 
    \begin{split} 
    \ket{\psi(t_{i+1} = t_i + \delta t)} & \\ 
    & \hspace*{-7em} = \ket{\psi(t_i)} + \frac{1}{6}\left(\ket{\psi_{i,1}} + 2 \ket{\psi_{i,2}} + 2 \ket{\psi_{i,3}} + \ket{\psi_{i,4}}\right), \label{eq:rg_propagation_scheme} 
    \end{split} 
\end{align}
where the intermediate increments are
\begin{subequations} 
    \begin{align} 
        \ket{\psi_{i,1}} &= -i \delta t\, V_I(t_i)\ket{\psi(t_i)}, \label{eq:psi_n1_rg_scheme} \\ 
        \ket{\psi_{i,2}} &= -i\delta t\, V_I\!\left(t_i + \frac{\delta t}{2}\right) \!\left(\ket{\psi(t_i)} + \frac{1}{2}\ket{\psi_{i,1}}\right), \label{eq:psi_n2_rg_scheme} \\ 
        \ket{\psi_{i,3}} &= -i\delta t\, V_I\!\left(t_i + \frac{\delta t}{2}\right) \!\left(\ket{\psi(t_i)} + \frac{1}{2}\ket{\psi_{i,2}}\right), \label{eq:psi_n3_rg_scheme}\\ 
        \ket{\psi_{i,4}} &= -i \delta t\, V_I(t_i + \delta t) \!\left(\ket{\psi(t_i)} + \ket{\psi_{i,3}}\right).
        \label{eq:psi_n4_rg_scheme}
    \end{align} 
\end{subequations}
We now express a generic update rule, $\ket{\psi^{\text{(new)}}} = V_I(t)\ket{\psi}$, in a compact form in order to implement this scheme in a numerically efficient manner. Recall that, in this work, a general state is decomposed as
\begin{align}
\ket{\psi} = \sum_{\alpha n, \beta m} c_{\alpha n, \beta m}\ket{\alpha \omega_n, \beta \omega_m, g} + \sum_{\alpha n} b_{\alpha n}\ket{\alpha \omega_n, e}.
\end{align}
Numerically, the state is represented as $\ket{\psi} \equiv \{C_{\mu \nu}, B_{\nu} \}$, where the compact index $\mu \equiv (\alpha,n)$ runs over the $2N_{\text{modes}}$ photonic excitations ($N_{\text{modes}}$ per branch). Because of bosonic commutation relations, the matrix $\bm{C}$ is symmetric, $C_{\mu\nu} = C_{\nu\mu}$. Defining the vector~\cite{pironRenormalizationTreatmentIR2026}
\begin{equation}
\bm{V}(t) = - i \sqrt{\frac{\gamma_0}{2L}}
\begin{pmatrix}
e^{i(\bm{\omega} - \omega_0)t} \\
e^{i(\bm{\omega} - \omega_0)t}
\end{pmatrix} \in \mathbb{C}^{2N_{\text{modes}} \times 1},
\end{equation}
where $\bm{\omega} = (\omega_n)_n$ denotes the discrete set of mode frequencies. The action of the interaction Hamiltonian on the state, $\ket{\psi^{\text{(new)}}} = V_I(t)\ket{\psi}$, can be written in compact form as
\begin{subequations}
\begin{align}
C^{\text{(new)}}_{\mu\nu} &= \frac{1}{2}\left(B_\mu V_\nu(t) + V_\mu(t) B_\nu\right), \\
B^{\text{(new)}}_\mu &= 2\sum_\nu C_{\mu\nu} V^*_\nu(t).
\end{align}
\end{subequations}
which can be vectorized in a straightforward manner as
\begin{align}
    \bm{C}^{\text{(new)}} = \frac{\bm{B} \bm{V}^T + \bm{V} \bm{B}^T}{2}, \quad \bm{B}^{\text{(new)}} = 2 \left(\bm{V}^{\dagger} \bm{C}\right)^T,
\end{align}
where we used the symmetry of $\bm{C}$. These expressions enable efficient computation of the increments, and thus a direct implementation of the scheme at the level of the coefficient tensors using \texttt{NumPy 1.26.2}.

\section{Two-photon kernels}
\label{appendix:kernels}

The equation of motion $\partial_t \ket{\psi} = -i V_I(t)\ket{\psi}$ can be treated within the Weisskopf–Wigner framework. Standard derivations can be found, e.g., in~\cite{scullyQuantumOptics1997}, and were followed in~\cite{pironRenormalizationTreatmentIR2026}. Extending this approach to the two-photon sector and integrating out the photonic amplitudes with initial condition $c_{\alpha n,\beta m}(0)=0$ yields the following effective equation for the atomic amplitudes:
\begin{align}
    \begin{split}
    \frac{d\, b_{\alpha n}}{dt} &= \int_0^t d\tau \,  K^{(1)}(\tau) b_{\alpha n}(t-\tau) \\
    & \hspace*{2em}+ \int_0^t d\tau \, \sum_{\beta m} K^{(2)}_{\alpha n, \beta m}(t, \tau) b_{\beta m}(t-\tau)
    \end{split}
\end{align}
where
\begin{align}
    \begin{split}
        K^{(1)}(\tau) &= -\frac{\gamma_0}{2L} \sum_{\beta m} e^{-i(\omega_m - \omega_0)\tau} \\
        K^{(2)}(t,\tau) &= -\frac{\gamma_0}{2L} e^{-i(\omega_m - \omega_n) t } e^{-i(\omega_m - \omega_0)(t - \tau)} 
    \end{split}
\end{align}
The first kernel $K^{(1)}$ coincides with the single-excitation kernel analyzed in~\cite{pironRenormalizationTreatmentIR2026}, while  $K^{(2)}$ captures a genuinely multi-photon contribution.

\end{document}